\documentclass[iop]{emulateapj}
\usepackage{graphicx,amssymb,amsmath}
\usepackage{hyperref,xcolor}
\usepackage{txfonts}
\usepackage{epstopdf}
\usepackage{natbib}
\usepackage{color}
\newcommand{\scs}{\scriptsize}

\shorttitle{Prospecting for chemical tags among open clusters}
\shortauthors{D. L. Lambert \& A. B. S. Reddy }

\begin{document} 


\title{Prospecting for chemical tags among open clusters}


\author{David L. Lambert}
\author{Arumalla B. S. Reddy}
\affil{W.J. McDonald Observatory and Department of Astronomy, The University of Texas at Austin, Austin, TX 78712-1205, USA}
\email{dll@astro.as.utexas.edu}




\begin{abstract}

Determinations of the chemical composition of red giants in a large sample of open clusters show that the abundances of the heavy elements La, Ce, Nd and Sm but not so obviously Y and Eu vary from one cluster to another across a sample all having about the solar metallicity. For La, Ce, Nd and Sm the amplitudes of the variations at solar metallicity scale approximately with the main $s$-process contribution to solar system material. Consideration of published abundances of field stars suggest that such a spread in heavy element abundances is present for the thin and thick disk stars of different metallicity. This new result provides an opportunity to chemically tag stars by their heavy elements and to reconstruct dissolved open clusters from the field star population.

\end{abstract}



\section{Introduction}
Open clusters (OCs) are valuable tools in probing the Galaxy's chemical evolution (GCE). They provide the opportunity to measure the  chemical composition of the Galaxy over distances of several kiloparsecs around the Sun and over a fraction of the Galactic disk's lifetime. Clusters and unbound stellar associations are born in dense molecular clouds. An open cluster is comprised of stars of different masses but with very similar, if not identical, initial chemical compositions. Over time, OCs dissolve and cluster members join the population of disk stars, already supplemented by stars from the unbound stellar associations \citep{LadaLada2003, PavBica2007}. The possibility exists that the majority of field stars now in the Galactic disk were born in dense molecular clouds and joined the disk after membership in an OC or an association. In this scenario, there is a close relationship between the compositions of OCs and the field star population. Composition differences between OCs of a similar generation will be echoed among field stars. Exploration of this idea requires accurate determinations of the compositions of OCs and field stars.

Here, we explore inter-cluster differences in chemical composition and, in particular, search for differences in elemental abundance ratios [El/Fe] among clusters of a common [Fe/H] from El=Na to Eu. Given that the majority of analyzed clusters have a near-solar metallicity ([Fe/H] $\sim 0.0$)  and belong to the thin disk, our investigation is centered on this particular collection of OCs.

Our exploration draws on our abundance analyses for 23 elements in 70 red giants belonging to 28 OCs with all analyses based on high-resolution optical spectra analyzed identically using LTE model atmospheres and a common line list \citep{Reddy2012,Reddy2013,Reddy2015,Reddy2016}. 
In short, we prepared a linelist of 300 absorption lines covering 23 elements from Na-Eu sampling all the major processes of stellar nucleosynthesis. The equivalent widths (EWs) were measured manually using the cursor commands in \textit{splot} package of {\scs IRAF} by fitting often a Gaussian profile and for a few lines a direct integration was performed for the best measure of EW. We performed a differential abundance analysis relative to the Sun by running the {\it abfind} driver of {\scs \bf MOOG}\citep{Sneden1973phd} adopting the 1D ATLAS9 stellar 
model atmospheres of \citet{CastelliKurucz2004}.

In most cases, the abundances are derived from the measured EWs but synthetic profiles were computed for lines affected by hyperfine structure (hfs) and isotopic splitting and/or affected by blends. The features analysed by spectrum synthesis included: Sc, Mn, Cu, Zn, Ba, and Eu. Our linelists have been tested extensively to reproduce the solar and Arcturus spectra before applying them to selected spectral features in the stellar spectra of program stars. We used a standard synthetic profile fitting procedure by running the {\it synth} driver of {\scs \bf MOOG} adopting the spectroscopically determined stellar parameters. Abundance differences are searched for among elements from all the major processes of stellar nucleosynthesis. 

The set of abundances for a given cluster may be considered to be the cluster's 'chemical tag', a term introduced by \citet{FreemanBland2002}. With the dissolution of clusters, the spread in chemical tags will be found among the field star population. Exploration of the chemical tagging technique may proceed in either a forward or a reverse direction. By the forward direction, we refer to the identification of members of a now dissolved OC from among a sample of field stars with well-determined chemical compositions.  Applications of the forward technique may be found in the papers by \citet{Ting2012}, \citet{Mitschang2014} and \citet{Hawkins2015}. Here, we refer to the reverse direction as the definition of inter-cluster abundance differences on whose existence successful exploitation of the forward direction obviously rests. As far as we are aware, the only previous application of a large sample of OCs in this reverse direction is by \citet{Blanco-Cuaresma2015}.

Successful reconstruction of dissolved OCs from the field star population depends on many factors but first and foremost must be the presence, magnitude and nature of inter-cluster chemical tags. Our analysis of cluster red giants provide an extensive set of compositions for OCs and, thus, a fine opportunity to address this key issue of chemical tags. Here, we present convincing evidence for very particular inter-cluster abundance differences suggested preliminarily earlier \citep{Reddy2015}.  Our analysis is comparable in scope and accuracy to that reported by \citet{Blanco-Cuaresma2015} except for one crucial  difference; our selection of elements but not that by \citet{Blanco-Cuaresma2015} includes several heavy elements --La, Ce, Nd, Sm and Eu -- which provide us with a novel insight into cluster-to-cluster differences in the contribution from the neutron-capture processes to the composition of the OCs. After defining these cluster-to-cluster differences, we searched for similar abundance differences among field dwarfs and giants.

\begin{figure*}
\begin{center}
\includegraphics[trim=0.1cm 9.6cm 2.6cm 4.6cm, clip=true,width=0.92\textwidth,height=0.26\textheight]{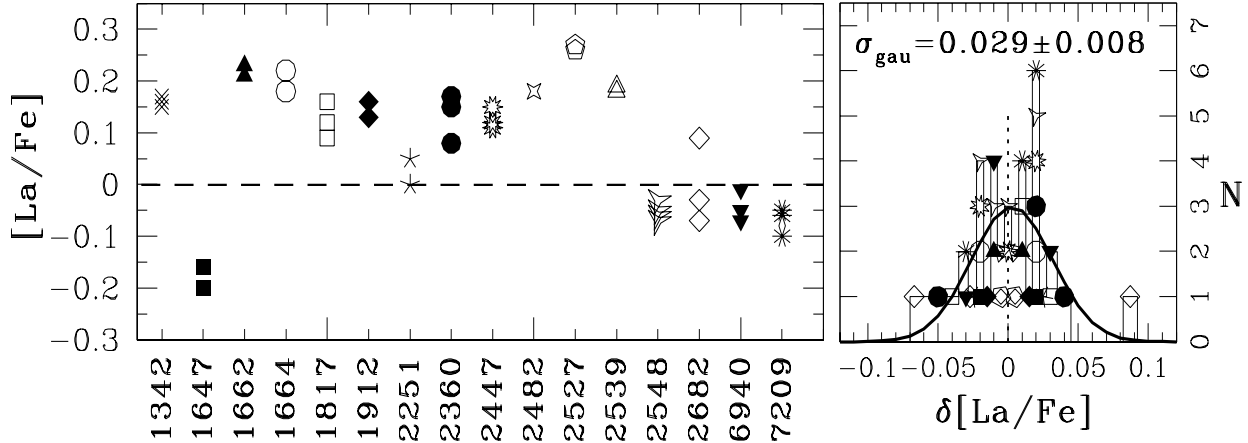}
\caption[]{Left panel: The relative abundances of the individual cluster members for the OCs in group i averaged over all available lines of the heavy $s$-process element La. Where the x-axis represent the NGC number of the cluster and the y-axis denote the abundance ratios for [La/Fe]. All the members within a cluster whose spectra are analysed for the chemical abundance analysis are denoted by the same symbol with each cluster represented by different symbols. Right panel: The distribution of residual abundance ratios of [La/Fe] ($\delta[La/Fe]$) for individual cluster members with each member represented by one unit on the y-axis. If two stars have the same magnitude of residual, then one star is displaced from the other by one unit vertically. All the symbols have their usual meaning as in the left panel. Where the residual abundance of [La/Fe] of a cluster member is obtained by subtracting its [La/Fe] from the average cluster abundance ratio of [La/Fe] (i.e., $\delta[La/Fe]$= [La/Fe]$_{star}$-$\langle[La/Fe]\rangle$$_{cluster}$). A Gaussian fit to the resultant histogram and the $\sigma_{\rm gau}$ is also shown. }
\label{inter_cluster}
\end{center}
\end{figure*}

This paper is organized as follows: We describe in Section 2 the sample of open clusters used to address the issue of chemical tags. Section 3 is devoted to discussing the cluster-to-cluster abundance variations among OCs and Section 4 to providing the convincing evidence for the abundance differences among field dwarfs and giants. Finally, we present our concluding remarks in Section 5.

\begin{figure*}
\begin{center}
\includegraphics[trim=0.05cm 10.9cm 0.5cm 4.1cm, clip=true,width=0.95\textwidth,height=0.18\textheight]{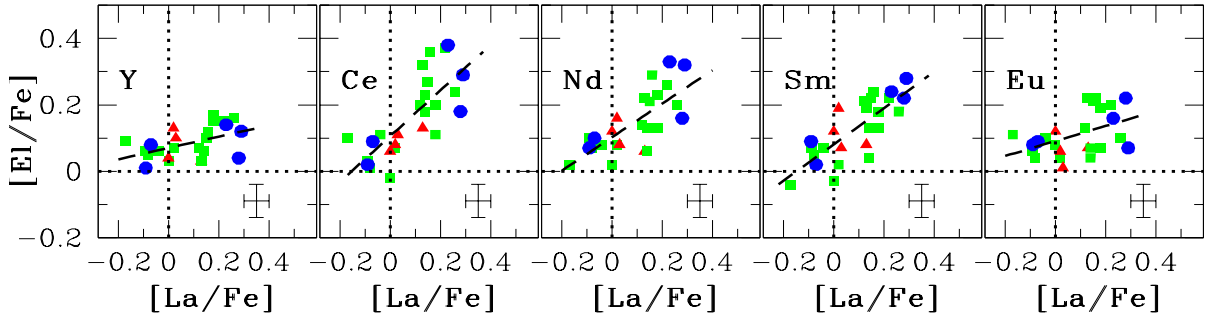}
\caption[]{The relative average cluster abundance ratios [La/Fe] vs [El/Fe] for elements (El) El$=$ Y, Ce, Nd,Sm and Eu. Clusters with mean [Fe/H] of $0.00\pm0.05$ (4 OCs), $-0.10\pm0.05$ (16 OCs) and $-0.20\pm0.05$ (5 OCs) are designated as red filled triangles, green filled squares and blue filled circles, respectively. The dashed line in each panel has the slope obtained from the least-squares fits with starting and end points set by the data points. 
}
\label{labatoeu_oc}
\end{center}
\end{figure*}

\section{The cluster sample}
Data on elemental abundances are taken from our analyses of red giant members of 28 OCs reported in \citet{Reddy2012,Reddy2013,Reddy2015,Reddy2016} with results for one to four stars per cluster and for usually 23 elements from Na to Eu. 
No attempt was made to combine our sample of OCs with the literature sample of OCs analysed by different authors whose analyses are subjected to systematic uncertainties of unknown magnitude from one another. The possible sources of systematic errors among the abundance analyses by different authors include differences in the signal-to-noise ratio (S/N) and resolution of the spectra, discrepancies in continuum tracing, adopted method of equivalent width measurements, differences in the model atmospheres, spectral-line selection, atomic data, abundance analyses technique (i.e., spectrum synthesis vs. EW analysis) and type of stars considered by different authors in their analysis. We direct the reader to \citet{Heiter2014} for a critical review of systematic errors. A simple merger of published abundances may mask subtle abundance trends. 
Moreover, a key enhancement of our analysis relative to analyses in the literature is that our OCs are essentially the only ones which sample well the elements from La to Eu.

This analysis draws on the selection of clusters belonging to the thin disk with particular attention focused on clusters around [Fe/H] of zero. Specifically, we consider in detail three groups: i) [Fe/H] = 0.0$\pm0.05$, ii) [Fe/H] $= -0.10\pm0.05$ and iii) [Fe/H] $= -0.20\pm0.05$. 
Group i consists of 4 clusters. Group ii is the largest group with 16 clusters and group iii is another sparse group with 5 clusters. With a single exception, these clusters have ages between about 100 Myr to 1 Gyr. Only three clusters, namely NGC 2345 ([Fe/H]$=-$0.26$\pm$0.03), NGC 2266 ([Fe/H]$=-$0.45$\pm$0.04), NGC 2632 ([Fe/H]$=+$0.08$\pm$0.03), are falling outside the [Fe/H] range defined by the OCs in groups i, ii, and iii and were excluded from the present analysis. All but NGC 2266 in our total OC sample belong to the thin disk population.

Red giants analyzed by us span small ranges in atmospheric parameters so that systematic errors across our sample are likely to be small. A majority of the red giants have an effective temperature of about 4900 K and a surface gravity of about 2.6 dex  with a cool limit at 4200 K and hot limit at 5400 K with a correlated range in surface gravity from 0.9 to 2.8 dex. Comparisons with the literature for main sequence and subgiant stars in clusters and the field will probably encounter offsets in  abundances [El/H] and abundance ratios [El/Fe] with several possible causes.
Two recent studies of dwarfs and giants in the same cluster have shown that abundance differences are small. \citet{Blanco-Cuaresma2015} analyzed five clusters and showed that abundance differences for [Fe/H] and [El/Fe] were less than 0.05 dex or almost every element considered except for Na which was enhanced in giants, an enhancement largely, if not wholly, arising from the first dredge-up affecting the giants. \citet{Dutra2016} provide a similar comparison for Fe in cool dwarfs and giants of the Hyades cluster. Abundance analyses of field dwarfs and giants conducted by different investigators may have larger systematic differences.

Although few stars were analyzed per cluster, when two or more stars were observed and analyzed consistent results were obtained. In particular, the paper's principal result -- that there are real inter-cluster differences for heavy elements with principal contributions from the $s$-process -- is supported by each of the few observed stars in  key clusters, i.e., our result is not an artifact arising from the overlooked inclusion of a Barium giant (a star whose $s$-process overabundances were contributed through mass-transfer from a companion). Additionally and significantly, the observed cluster members are at luminosities where internal $s$-process enrichment is not anticipated. 

We probe through the Figure \ref{inter_cluster} the presence of internal chemical homogeneity among members of given cluster and the significant cluster-to-cluster abundance spread in [El/Fe] for the heavy $s$-process elements among clusters of a common [Fe/H]. Figure \ref{inter_cluster} shows the [La/Fe] values measured for individual cluster members for the OCs in group i. Examination of the Figure \ref{inter_cluster} suggests that the typical star-to-star abundance dispersion in [La/Fe] as quantified by fitting a Gaussian function to the residual abundance of [La/Fe] of a star is about 0.03$\pm$0.01 dex (see the figure caption). The OCs in Figure \ref{inter_cluster} demonstrate  that the internal chemical homogeneity is very small relative to the cluster-to-cluster inhomogeneity which for [La/Fe] is 0.4 dex.

The mean [La/Fe] values for the clusters in Figure \ref{inter_cluster} range over 0.4 dex (i.e., $-$0.17 dex for NGC 1647 to $+$0.26 dex for NGC 2527); this range is almost ten times larger than the internal abundance dispersion found for [La/Fe] from a given cluster. Inspection of the [X/Fe] for the other heavy elements confirms that the cluster-to-cluster abundance spread seen in [La/Fe] is repeated for Ce, Nd and Sm for which we found the typical star-to-star abundance dispersion of 0.027$\pm$0.008 dex, 0.032$\pm$0.014 dex and 0.035$\pm$0.007 dex, respectively. 
In contrast to the heavy $s$-process elemental abundances, the cluster-to-cluster scatter is small for the light $s$-process elements -- Y (Figure \ref{labatoeu_oc}) and Zr. Therefore, the striking spread in the abundances of heavy $s$-process elements from one cluster to another whose members exhibit high level of chemical homogeneity within a given cluster confirms that there is real inter-cluster differences in [El/Fe] for La to Sm.

\section{Abundance differences between clusters} \label{abu_diff:clusters}
\subsection{Heavy elements}
Consideration of the red giants' compositions led \citet{Reddy2015} to suggest that there was a spread in the abundances of heavy elements La, Ce and Nd  which exceeded plausible estimates of the measurement errors and that  presence of a spread did not extend to the lighter element Y. With our now larger sample of OCs, these preliminary results are confirmed and extended. Figure \ref{labatoeu_oc} constructed from our sample of OCs highlight  the key aspects of the  intra-cluster  abundance differences among the heavier elements. 

Figure \ref{labatoeu_oc} for OCs from [Fe/H] groups i, ii and iii shows that the average [El/Fe] for each cluster is well correlated for La, Ce, Nd and Sm but Y and Eu show a very much weaker correlation with [La/Fe]. For the strong Ba\,{\sc ii} 5853 \AA\ and 6496 \AA\ lines, the larger abundance uncertainty arising principally from the sensitivity to the microturbulence likely masks the abundance spread.

The spread in [El/Fe] for La to Sm is a real effect. The small offsets -- i.e., [Ce/Fe] $\simeq +0.05$ at [La/Fe] $ = 0.0$ -- may not be real because of various systematic effects including the adopted $gf$-values and non-LTE effects. Differences in the behavior of La, Ce, Nd and Sm  (also Ba) are approximately correlated with the $s$-process contributions made to the solar system abundances for these elements. Synthesis of these elements is provided by both the neutron capture $s$- and $r$-processes. According to \citet{Burris2000}, the $s$-process contribution to solar system abundances is 85\% for Ba, 75\% for La, 81\% for Ce, 47\% for Nd, 34\% for Sm and only 3\% for Eu. This same reference assigns 72\% of Y and 81\% of Zr to the $s$-process but, as Figure \ref{labatoeu_oc} indicates, the spread seen in La and Ce, for example, is not repeated for Y.

The pattern of decreasing heavy element enrichment in the sequence La to Eu resembles but does not exactly match that of the solar system $s$-process. Linear least-squares fits to the data in Figure \ref{labatoeu_oc} give slopes for [El/Fe] versus [La/Fe] of 0.17$\pm0.06$ (R$=$0.52) for Y, 0.69$\pm0.12$ (R$=$0.79) for Ce, 0.50$\pm0.09$ (R$=$0.69) for Nd, 0.54$\pm0.09$ (R$=$0.71) for Sm and $0.22\pm0.09$ (R$=$0.49) for Eu, where R is the Pearson product-moment correlation coefficient. In particular, the Eu trend is not compatible with the 3$\%$ expected from the solar system's $s$-process. 
A simple interpretation of Figure \ref{labatoeu_oc} is that the  main $s$-process contributions influencing the spread in Ba-Sm abundances  contribute little to the spread in Y whose abundance is most likely then set primarily by the weak $s$-process occurring during He and C shell hydrostatic burning in massive stars. The source of the main $s$-process is, of course, thermally-pulsing AGB stars. 

\begin{figure}
\begin{center}
\includegraphics[trim=0.1cm 10.4cm 12.8cm 4.4cm, clip=true,width=0.48\textwidth,height=0.24\textheight]{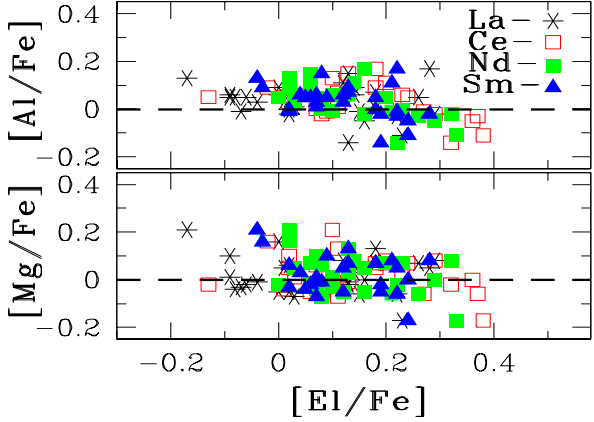}
\caption[]{The runs of [Al/Fe] and [Mg/Fe] with the [El/Fe] for El= La, Ce, Nd and Sm for our sample of 27 thin disk OCs.}
\label{almg_heavy}
\end{center}
\end{figure}

Variations from cluster-to-cluster exhibited in Figure \ref{labatoeu_oc} are at the precision provided by our abundance analyses of red giants uncorrelated with abundances of Fe-group and lighter elements. Figure \ref{almg_heavy} shows the La$-$Sm abundances plotted with the Al and Mg abundances. The large spread in [El/Fe] for La$-$Sm is in sharp contrast with the much smaller spread in [Al/Fe] and [Mg/Fe]. There may be a tendency for [Al/Fe] and, perhaps, also[ Mg/Fe] to decline slightly with increasing [El/Fe] for La$-$Sm.

As a supplement to visual inspection for picking out abundance differences among stars of similar metallicities, we compare the distributions of [El/H] and [El/Fe] with those expected from the measurement uncertainties and the assumption that all stars have of the same [Fe/H] have identical [El/Fe]. We quantify the cluster-to-cluster scatter in the elemental abundance ratios, [El/H] and [El/Fe] for an element El by fitting a Gaussian function to histograms of number of stars (y-axis) versus the average abundance ratios (x-axis) [El/H] and [El/Fe] of an element El measured for red giants in all OCs in groups i and ii covering the narrow [Fe/H] interval of 0.0$\pm$0.05 dex to $-$0.10$\pm$0.05 dex.

\begin{table} 
\centering 
 {\fontsize{7.5}{8}\selectfont 
\caption{Estimates of $\sigma_{\rm gau}$s from Gaussian fits to the histogram of number of stars versus the average abundance of an element El measured for red giants in all the OCs in groups i and ii along with the measurement error expected for the same element El due to uncertainties in the stellar atmospheric parameters and the equivalent width measurements. Numbers in the parentheses indicate the number of clusters used for the estimate of $\sigma_{\rm gau}$. }  
\vspace{0.1cm}
\label{sigmas_oc}
\begin{tabular}{cll|cc}   \hline
\multicolumn{1}{c}{El} & \multicolumn{1}{c}{$\sigma_{\rm gau}$ ([El/H])} & \multicolumn{1}{c}{$\sigma_{\rm [El/H]_{exp.}}\pm$std} & \multicolumn{1}{c}{$\sigma_{\rm gau}$ ([El/Fe])} & \multicolumn{1}{c}{$\sigma_{\rm [El/Fe]_{exp.}}\pm$std}   \\ \hline

  Na (20)           & $0.162\pm0.046$   & $0.083\pm0.006$   & $0.108\pm0.051$   & $0.047\pm0.006$   \\
  Mg (20)           & $0.117\pm0.051$   & $0.060\pm0.020$   & $0.102\pm0.035$   & $0.057\pm0.015$   \\
  Al (20)           & $0.156\pm0.062$   & $0.050\pm0.010$   & $0.091\pm0.028$   & $0.057\pm0.012$   \\
  Si (20)           & $0.122\pm0.041$   & $0.067\pm0.012$   & $0.062\pm0.019$   & $0.113\pm0.025$   \\
  Ca (20)           & $0.106\pm0.031$   & $0.107\pm0.021$   & $0.091\pm0.035$   & $0.053\pm0.015$   \\
  Sc (20)           & $0.081\pm0.031$   & $0.127\pm0.021$   & $0.063\pm0.017$   & $0.070\pm0.026$   \\
  Ti (20)           & $0.103\pm0.041$   & $0.123\pm0.015$   & $0.061\pm0.018$   & $0.063\pm0.025$   \\
  V  (20)           & $0.135\pm0.054$   & $0.147\pm0.029$   & $0.078\pm0.019$   & $0.083\pm0.032$   \\
  Cr (20)           & $0.079\pm0.017$   & $0.107\pm0.012$   & $0.038\pm0.009$   & $0.050\pm0.020$   \\
  Mn (20)           & $0.142\pm0.048$   & $0.097\pm0.006$   & $0.091\pm0.028$   & $0.043\pm0.015$   \\
  Fe (20)           & $0.039\pm0.008$   & $0.093\pm0.010$   &    $\ldots$       &    $\ldots$       \\
  Co (20)           & $0.139\pm0.034$   & $0.080\pm0.008$   & $0.073\pm0.018$   & $0.057\pm0.012$   \\
  Ni (20)           & $0.033\pm0.007$   & $0.073\pm0.006$   & $0.031\pm0.009$   & $0.070\pm0.026$   \\
  Cu (20)           & $0.194\pm0.066$   & $0.077\pm0.015$   & $0.112\pm0.029$   & $0.053\pm0.023$   \\
  Zn (20)           & $0.170\pm0.045$   & $0.087\pm0.006$   & $0.155\pm0.040$   & $0.113\pm0.006$   \\
  Y  (20)           & $0.086\pm0.033$   & $0.063\pm0.012$   & $0.064\pm0.020$   & $0.090\pm0.010$   \\
  Zr (20)           & $0.187\pm0.046$   & $0.143\pm0.031$   & $0.134\pm0.049$   & $0.087\pm0.035$   \\
  Ba (20)           & $0.152\pm0.031$   & $0.103\pm0.006$   & $0.178\pm0.042$   & $0.080\pm0.010$   \\
  La (20)           & $0.158\pm0.038$   & $0.060\pm0.010$   & $0.150\pm0.047$   & $0.073\pm0.006$   \\
  Ce (20)           & $0.201\pm0.049$   & $0.067\pm0.006$   & $0.196\pm0.051$   & $0.083\pm0.006$   \\
  Nd (20)           & $0.139\pm0.047$   & $0.063\pm0.006$   & $0.112\pm0.030$   & $0.077\pm0.006$   \\
  Sm (20)           & $0.123\pm0.039$   & $0.070\pm0.010$   & $0.113\pm0.033$   & $0.077\pm0.012$   \\
  Eu (20)           & $0.070\pm0.017$   & $0.053\pm0.006$   & $0.071\pm0.017$   & $0.083\pm0.006$   \\

\hline
\end{tabular}
 }
\end{table}

Entries in Table \ref{sigmas_oc} provide the estimates of the $\sigma_{\rm gau}$ of the Gaussian for [El/H] and [El/Fe] for all elements along with the measurement errors ($\sigma_{exp}$) expected due to uncertainties in the stellar atmospheric parameters and the equivalent width  measurements. The average of individual $\sigma_{\rm total}$ derived for three representative stars spanning the T$_{\rm eff}$ and log~$g$ range of the entire sample of red giants analyzed in OCs is adopted as a measure of the $\sigma_{exp}$. Where the $\sigma_{\rm total}$ for each of the elements El is computed by adding quadratically the uncertainties in the equivalent width  measurements and the changes in abundance ratios [El/H] and [El/Fe] caused by varying the estimated stellar parameters of a star by an amount equal to typical errors of $\pm$100 K in $T_{\rm eff}$, $\pm$0.1 cm s$^{-2}$ in log~$g$ and $\pm$0.1 km s$^{-1}$ in $\xi_{t}$. As summarized in Table \ref{sigmas_oc}, the $\sigma_{\rm gau}$ of the Gaussian and the $\sigma_{exp}$ varies from element to element. 

Entries in Table \ref{sigmas_oc} for $\sigma_{\rm gau}$ are well matched to values of $\sigma_{exp}$ for all but a few elements. Obvious exceptions include Na, Mg, Al, Mn, Cu, and the heavy $s$-process elements Ba, La, Ce, Nd and Sm for which the $\sigma_{\rm gau}$ exceeds by almost a factor of two the estimated measurement error $\sigma_{exp}$. These heavy element -- exceptions are not regarded as artifacts since the abundance of these elements are analyzed from  well measured equivalent widths and the analysis takes into account, as necessary, the hyperfine structure and isotopic shifts in the spectrum synthesis of lines, and we also stress that the usage of same line list, atmospheric models and the reference solar abundances results in tight control of systematic errors. Additionally, the Y and Eu abundances show no excess spread beyond that expected from the measurement uncertainties. 

A noticeable feature of Table \ref{sigmas_oc} is that observed $\sigma_{gau}$ for both [El/H] and [El/Fe] for the sequence Ba to Eu scale quite closely with the main $s$-process contribution to solar system material: the $\sigma$ for Ba, La and Ce are comparable with Nd and Sm providing smaller values and Eu has the smallest $\sigma$. With the sole exception of Na whose presence in giants can be affected by mild Na enrichment as a result of the first dredge-up \citep{Smiljanic2012,KarakasLattanzio2014}, the intrinsic cluster-to-cluster abundance scatter in [El/H] and [El/Fe] is presumed to be responsible for the detectably larger than expected sigmas for Mg, Al, the heavy $s$-process elements and apparently Cu and Zn. The Na 
abundances will be discussed elsewhere.

\subsection{Previous studies of heavy elements}
Published abundance analyses of OCs have not focussed sharply on the possibility of intrinsic scatter among the OC population in heavy element abundances. Attention has generally been devoted to determinations of the abundance gradient and the abundance-age relations as obtained from different elements. Additionally, recent investigations have tended to give priority to investigating the claim by \citet{Dorazi2009} that the  barium abundance increases sharply with decreasing age of an OCs: [Ba/Fe] increases from 0.0 at about 4 Gyr to $+0.6$ in very young clusters. 

Perhaps, the first acknowledgment of heavy element scatter was provided by \citet{Maiorca2011} who determined Y and Ce abundances for an OC
sample with much overlap with that considered by D'Orazi et al. Maiorca et al. noted that three OCs showed high [Ce/Fe] ratios but `this dispersion'
was not assigned an intrinsic origin but suspected to arise from a systematic error in the analysis. (Neither Zr nor La were measured
for these three OCs.)

\citet{JacobsonFriel2013} measured heavy elements Zr, Ba, La and Eu and lighter elements in 19 OCs with ages from 700 Myr to 10 Gyr and extending from Galactocentric distances of 8.4 to 17.7 kpc with an emphasis on abundance variations with age and Galactocentric distance. Their abundances of Zr, La and Ba were correlated with the [La/Fe] spread, similar to that in our Figure \ref{labatoeu_oc}. The correlation between La and Eu was weak.
In this examination, clusters were divided into young and old with the dividing line at 4 Gyr. All but one of our group i, ii and iii clusters would be in the young category. No distinction was made by Galactocentric distance. 

\begin{figure*}
\begin{center}
\includegraphics[trim=0.05cm 10.9cm 0.5cm 4.1cm, clip=true,width=1.1\textwidth,height=0.20\textheight]{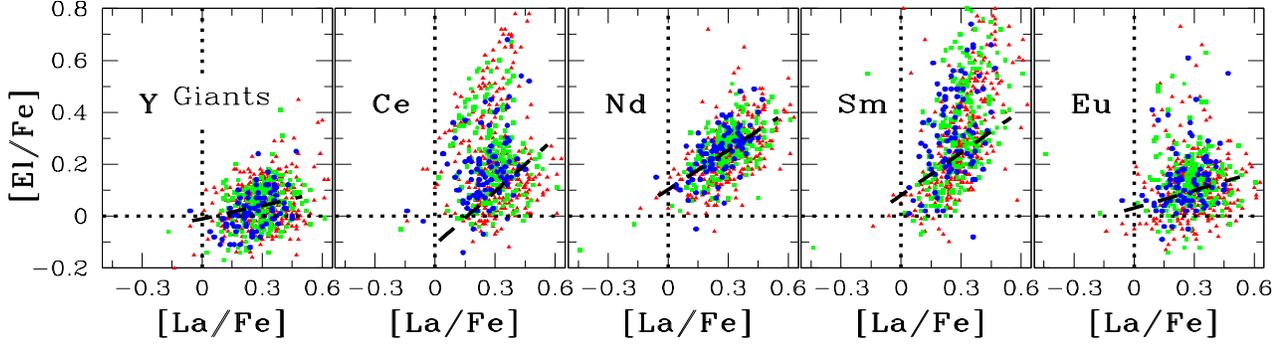}
\caption[]{Same as figure \ref{labatoeu_oc}, but for a sample of the local F, G and K giants from \citet{Luck2015} for [Fe/H] groups i, ii and iii. The dashed line has same length and slope of the corresponding line in Figure \ref{labatoeu_oc}. }
\label{labatoeu_luck}
\end{center}
\end{figure*}

\begin{figure*}
\begin{center}
\includegraphics[trim=0.05cm 11.4cm 7.5cm 4.5cm, clip=true,width=0.85\textwidth,height=0.15\textheight]{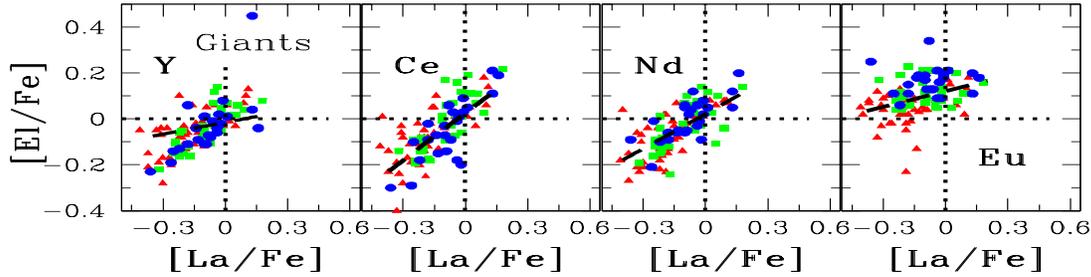}
\caption[]{Same as figure \ref{labatoeu_oc}, but for a sample of giants from \citet{Mishenina2006,Mishenina2007} for [Fe/H] groups i, ii and iii. The dashed line has same length and slope of the corresponding line in Figure \ref{labatoeu_oc}. }
\label{labatoeu_mishenina}
\end{center}
\end{figure*}

\citet{Mishenina2014,Mishenina2015} considered 13 OCs with a similar age and Galactocentric distance spread as Jacobson \& Friel. Plots of [El/Fe] for El = Y, Ba and La versus age and [Fe/H] showed a large dispersion for Ba, a $\pm0.2$ dispersion for La and a smaller dispersion for Y, estimates compatible with Figure \ref{labatoeu_oc}. But these plots also connect measurements for a cluster by different investigators and in almost every case the difference between measurements exceeds the apparent dispersion. This disquieting result highlights the need, as in our study, for a uniform analysis of a large sample of OCs.

Curiously, recent analyses of heavy element abundances for main sequence stars in stellar associations show no spread in abundances. Dwarf stars from a total of six associations have now been analysed \citep{Biazzo2012,Dorazi2012,Reddy2015} for elements from Li to Eu. The associations have [Fe/H] $\simeq 0.0$ and [El/Fe] $\simeq 0.0$ for all El except Li and Ba. Lithium is not at all a surprising exception. Barium provides [Ba/Fe] $\simeq +0.2$ when other heavy elements Y to Eu give [El/Fe] $= 0.0$ with a small scatter. 
This lack of scatter among dwarf stars in young associations (age less than 200 Myr) in the solar neighborhood is at odds with the obvious scatter (Figure \ref{labatoeu_oc}) among red giants from OCs with age of 100 to 1000 Myr at distances of up to 2000 pc. What astrophysical explanation can there be for this difference between stellar assemblies born from molecular clouds?  Or is this an insidious result of a small sample size or are systematic errors raising their head here?

\section{Abundance differences among field stars}
The idea that dispersal of OCs and stellar associations may control Galactic field star populations implies identical field and cluster stars of comparable age should have very similar compositions. Most of the dispersed stars lose memory of their spatial location and kinematics and eventually end up in very different parts of phase space in the Galaxy, but their natal chemistry, except for the elements affected by stellar evolutionary changes, is thought to be preserved in stellar photospheres throughout their lifetime. Therefore, the chemical composition is the tag or label identifying stars of common origin even after the spatial information of a cluster's birthplace has vanished. Then, one expects a close, if not exact, correspondence between the compositions of OCs and field stars. This expectation was searched for in \citet{Reddy2015} through a global comparison of the runs of [El/Fe] with [Fe/H] and age for field stars and OCs drawn from our previous studies and from the literature where Reddy et al. concluded that the mean values of [El/Fe] and their $\sigma$ except for heavy elements were similar for field and cluster stars over the [Fe/H] range sampled by the OCs; [El/Fe] for the latter elements span a range exceeding that expected purely from measurement uncertainties. 

With the exception of NGC 2682 (M 67) and NGC 752 with an age of 4.1 Gyr and 1.1 Gyr, respectively, our clusters with [Fe/H]$=0.0\pm0.05$ (4 OCs) and $-0.1\pm0.05$ (16 OCs) have similar young ages (0.07-0.6 Gyr) and may not have moved far away radially from their birthplaces. All these clusters occupy a limited range in Galactocentric distance (7.7 to 9.9 kpc). Hence, they offer an excellent opportunity to explore chemical inhomogeneities in the Galactic disk at the OCs' birthplaces. Stars in our sample are similar red giants analyzed identically and, thus, systematic errors affecting the abundances and abundance ratios [El/Fe] should be consistent (and small) across the sample which spans a small range in metallicity [Fe/H].

The OCs analyzed for the inter-cluster abundance differences are presumably representative sample of clusters approaching dissolution. Possibly, the OCs which have completed dissolution were similar. Then, among field stars of similar [Fe/H] one might expect to find examples  with enrichments of up to a factor of two for  Ba, La, Ce and Nd but showing no enrichment of light $s$-process elements (Sr, Y and Zr) and only a weak enrichment of Eu. 

A successful search for such stars would demonstrate the transference to field stars of a chemical tag held by an OC to the field star population. We study several samples of field stars: local giants analyzed by \citet{Luck2015}, \citet{Mishenina2006} and \citet{Mishenina2007} and thin disk dwarfs with heavy element abundances determined by \citet{BatBensby2016} and other elements determined previously by \citet{Bensby2014}.

\begin{figure*}
\begin{center}
\includegraphics[trim=0.05cm 11.7cm 5.8cm 4.1cm, clip=true,width=0.90\textwidth,height=0.15\textheight]{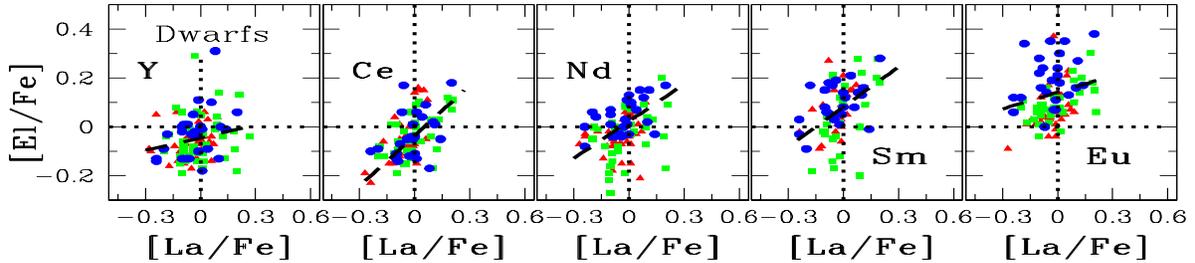}
\caption[]{Same as figure \ref{labatoeu_oc}, but for a sample of dwarfs from \citet{BatBensby2016} for [Fe/H] groups i, ii and iii. The dashed line has same length and slope of the corresponding line in Figure \ref{labatoeu_oc}. }
\label{labatoeu_bensby}
\end{center}
\end{figure*}

\citet{Luck2015} analyzed 1133 local F, G and K giants for a suite of elements including Y, Zr, Ba, Ce, Pr, Nd and Eu.  We scanned this sample for the equivalent of our group i, ii and iii stars. For the metallicities spanned by the three groups, Figure \ref{labatoeu_luck} shows the result which is to be compared with Figure \ref{labatoeu_oc} giving results for individual OCs. The dashed line in a given panel of Figure \ref{labatoeu_luck} has the slope and length of the corresponding dashed line in Figure \ref{labatoeu_oc} but minor adjustments in positioning of a line were made in recognition of the possibility of small offsets in abundances between the two analyses. Several similarities between field and cluster stars are apparent. Abundances of Y and Eu  are uncorrelated with the La abundance for both field and cluster stars. Abundances of La, Ce, Nd and Sm are correlated for field giants but not as tidily for the clusters. For these field giants, the cleanest correlation is between La and Nd with amplitudes for both elements similar to the amplitudes seen for the clusters. For Ce and Sm, the sample shows much larger amplitudes in [El/Fe] which may be due unrecognized blends affecting some of the Ce\,{\sc ii} and Sm\,{\sc ii} lines.

A much smaller sample of giants with coverage of the heavy elements is provided by \citet{Mishenina2006} and \citet{Mishenina2007}. Heavy element abundances among groups i, ii and iii are plotted in Figure \ref{labatoeu_mishenina}. La, Ce and Nd abundances are correlated but in contrast to Figures \ref{labatoeu_oc} and \ref{labatoeu_luck} Y also appears to be correlated with the heavy $s$-process elements but this correlation is weak for stars within each group. Amplitudes of the spreads for La, Ce and Nd are consistent with the spread for the OCs (see the dashed line in each panel of Figure \ref{labatoeu_mishenina}). Appearance of a correlation between Y and heavy $s$-process elements runs counter to our results. Systematic errors in the [Fe/H] abundances would introduce a spurious correlation of unit slope between two elements with no intrinsic correlation but since [Eu/Fe] and [La/Fe] are uncorrelated this simple explanation for the apparent [Y/Fe] - [La/Fe] correlation is given low weight. Although, in contrast to Figure \ref{labatoeu_luck}, considerable offsets are needed to match abundances in Figure \ref{labatoeu_mishenina} to those in Figure \ref{labatoeu_oc}, the slope and end-points of dashed lines from Figure \ref{labatoeu_oc} are fine fits to the distribution of the abundances in Figure \ref{labatoeu_mishenina}. On the reasonable assumption that the offsets arise from a difference in systematic errors between the analyses, it may be claimed that Mishenina et al.'s sample of field giants display a very similar spread in heavy element abundances to our sample of OCs. 

\begin{table} 
\centering 
{\fontsize{8}{5}\selectfont 
\caption{The $\sigma$ estimate from the Gaussian fit ($\sigma_{\rm gau}$) to the chemical abundance ratios of the field dwarfs (\cite{Bensby2014, BatBensby2016}) and field giants (Luck 2015) covering the same metallicity range as that of the OCs in this study (i.e. [Fe/H]$=$0.00$\pm$0.05 to $-$0.10$\pm$0.05 dex)  }  \vspace{0.1cm}
\label{sigmas_battistini}
\begin{tabular}{ccc|cc}   \hline
 \multicolumn{1}{c}{ } & \multicolumn{2}{c}{Dwarfs (120 stars)} \vline & \multicolumn{2}{c}{Giants (550 stars)}  \\ \cline{1-5}
 \multicolumn{1}{c}{El} & \multicolumn{1}{c}{$\sigma_{\rm gau}$ ([El/H])} & \multicolumn{1}{c}{$\sigma_{\rm gau}$ ([El/Fe])} \vline  &
 \multicolumn{1}{c}{$\sigma_{\rm gau}$ ([El/H])} & \multicolumn{1}{c}{$\sigma_{\rm gau}$ ([El/Fe])} \\ \hline

  Na    & $0.104\pm0.013$   & $0.064\pm0.010$   & $0.141\pm0.008$   & $0.096\pm0.006$   \\
  Mg    & $0.089\pm0.011$   & $0.072\pm0.008$   & $0.096\pm0.007$   & $0.065\pm0.004$   \\
  Al    & $0.090\pm0.012$   & $0.096\pm0.023$   & $0.084\pm0.005$   & $0.053\pm0.003$   \\
  Si    & $0.085\pm0.012$   & $0.031\pm0.004$   & $0.088\pm0.005$   & $0.062\pm0.003$   \\
  Ca    & $0.075\pm0.011$   & $0.028\pm0.003$   & $0.092\pm0.009$   & $0.033\pm0.001$   \\
  Sc    &    $\ldots$       &    $\ldots$       & $0.069\pm0.005$   & $0.058\pm0.003$   \\
  Ti    & $0.087\pm0.011$   & $0.038\pm0.006$   & $0.092\pm0.009$   & $0.037\pm0.002$   \\
  V     &    $\ldots$       &    $\ldots$       & $0.127\pm0.011$   & $0.069\pm0.003$   \\
  Cr    & $0.087\pm0.014$   & $0.023\pm0.002$   & $0.099\pm0.010$   & $0.027\pm0.001$   \\
  Mn    &    $\ldots$       &    $\ldots$       & $0.128\pm0.014$   & $0.060\pm0.003$   \\
  Fe    & $0.087\pm0.014$   &    $\ldots$       & $0.077\pm0.035$   &    $\ldots$       \\
  Co    &    $\ldots$       &    $\ldots$       & $0.078\pm0.007$   & $0.042\pm0.003$   \\
  Ni    & $0.088\pm0.012$   & $0.032\pm0.004$   & $0.103\pm0.013$   & $0.047\pm0.005$   \\
  Cu    &    $\ldots$       &    $\ldots$       & $0.219\pm0.015$   & $0.188\pm0.011$   \\
  Zn    & $0.136\pm0.019$   & $0.089\pm0.013$   & $0.349\pm0.027$   & $0.366\pm0.031$   \\
  Y     & $0.096\pm0.016$   & $0.082\pm0.0101$  & $0.110\pm0.009$   & $0.079\pm0.004$   \\
  Zr    & $0.103\pm0.014$   & $0.113\pm0.018$   & $0.185\pm0.012$   & $0.163\pm0.009$   \\
  Ba    & $0.122\pm0.017$   & $0.039\pm0.005$   & $0.129\pm0.007$   & $0.141\pm0.009$   \\
  La    & $0.170\pm0.049$   & $0.148\pm0.029$   & $0.130\pm0.008$   & $0.126\pm0.009$   \\
  Ce    & $0.170\pm0.027$   & $0.164\pm0.036$   & $0.135\pm0.009$   & $0.139\pm0.008$   \\
  Nd    & $0.134\pm0.023$   & $0.102\pm0.015$   & $0.091\pm0.004$   & $0.100\pm0.007$   \\
  Sm    & $0.157\pm0.018$   & $0.184\pm0.039$   & $0.226\pm0.016$   & $0.231\pm0.017$   \\
  Eu    & $0.089\pm0.009$   & $0.072\pm0.009$   & $0.102\pm0.007$   & $0.079\pm0.004$   \\

\hline
\end{tabular}  }
\end{table}

For thin disk dwarfs, we consider the large sample analyzed by Bensby et al. (2014) with heavy element abundances provided by Battistini \& Bensby (2016). Abundances for [Fe/H] groups i, ii and iii are presented in Figure \ref{labatoeu_bensby} with the dashed lines taken from Figure \ref{labatoeu_oc} but with offsets adjusted to get the best fit. Although the dashed lines provide a fair account of the abundances for the dwarfs, the account is not as convincing as that offered by the giants and Figure \ref{labatoeu_mishenina}. 
Comparison of the figures for field giants and dwarfs suggests that the scatter about the dashed line in each panel is larger for dwarfs than for giants with a contribution for dwarfs resulting from systematic differences between [Fe/H] groups i, ii and iii. Systematic shifts between groups are especially noticeable for Nd, Sm and Eu, the three elements with smaller contributions from a main $s$-process than La and Ce. Y does not exhibit this scatter. An offset between groups is especially noticeable for Eu. The ages of the thin disk dwarfs (2 to 10 Gyr) are considerably greater than for our OCs (0.1 to 1 Gyr). This age difference and Figure \ref{labatoeu_mishenina} and \ref{labatoeu_bensby} may suggest that the nature of a chemical tag for open clusters evolves over the Milky Way's lifetime. Possibly, this particular piece of evidence refers to contributions from the $r$-process sources (Type II supernovae or merging neutron stars?).

For an alternative consideration of the abundances across groups i ii and iii  for both Luck's (2015)  sample of giants and
Bensby et al.'s (2014) and Battistini \& Besby et al.'s (2016) sample of thin disk dwarfs, we present in Table \ref{sigmas_battistini} $\sigma$ from the distributions of [El/H] and [El/Fe]. This table shows that the majority of elements lighter than Ni have very similar and small $\sigma$s, e.g., $\sigma \simeq 0.09$ for [El/H] and $\simeq 0.06$ for [El/Fe] for the dwarfs and giants. Y and Eu show a $\sigma$ of a similar value for dwarfs and only a slightly larger value for the giants. Our OCs sample (Table \ref{sigmas_oc}) show $\sigma$ for Y and Eu similar to these values. 
The reported $\sigma$s are similar to the expected values and, thus, the intrinsic contributions to [El/H] and [El/Fe] are small. For heavier elements, the $\sigma$s from Ba to Sm are larger than expected and, thus, there is a real star-to-star scatter. In the case of Ba to Eu, $\sigma$s for Ce and Sm for giants may be ignored (see Figure \ref{labatoeu_luck}) but values for Ba, La, Nd and Eu are close to the values (Table \ref{sigmas_oc}) for OCs. For the field dwarfs, the $\sigma$s for Y to Eu are generally compatible with values for the OCs. Table \ref{sigmas_battistini} is, of course, a reflection of Figures \ref{labatoeu_luck} and \ref{labatoeu_mishenina} and their relation to Figure \ref{labatoeu_oc}.

Presence of abundance differences based on heavy element [El/Fe] for Ba to Sm has been established for OCs in the thin disk with near solar metallicity (groups i, ii and iii) and identified as present among similar metallicity field giants and dwarfs. Thus, there is a possibility that such differences may serve as a chemical tag. It is of interest to know if the tag may be found among more metal-poor thin disk stars and stars in the thick disk. To investigate these points, we separated the thin and thick disk field dwarfs analysed by \citet{Bensby2014} and \citet{BatBensby2016} into several metallicity bins. 

For the thin disk, the available sample extends to about a [Fe/H] of $-0.30$, the $\sigma$s for Ni and lighter elements remain small, i.e., higher abundance precision is needed to extract the intrinsic scatter in [El/Fe]. Al is a possible exception with the $\sigma$ for [Al/Fe] increasing with decreasing [Fe/H]. For heavy elements, the $\sigma$ for La to Eu, as [Fe/H] declines, depart from the clean pattern in Table \ref{sigmas_battistini} (also, that in Table \ref{sigmas_oc}) and at [Fe/H] $\simeq -0.25$ the $\sigma$s for [El/H] for La, Ce, Nd, Sm and Eu are 0.14, 0.24, 0.12, 0.18 and 0.16, respectively, and about 40 stars contributing. 
 
For the thick disk which is well sampled to about a [Fe/H] of $-0.90$, the abundance pattern of the chemical tag from the OCs is not perfectly repeated. In particular Sm has a larger spread than Nd and Eu has a large spread. This pattern resembles that found for the low [Fe/H] end of the thin disk.

\section{Concluding remarks}
Inspired by a search for chemical tags among open clusters, our abundance analysis using high-resolution optical spectra of a representative sample of OCs led to the first definition of a cluster-to-cluster difference in the abundances of heavy elements. The pattern of these differences is summarized well by Figure \ref{labatoeu_oc} for clusters with metallicities [Fe/H] $\sim 0.0$: inter-cluster differences are small for Y and Eu and larger for La, Ce, Nd and Sm with amplitudes that scale quite well with the neutron capture main $s$-process's contribution to solar system material. A small  cluster-to-cluster difference for Y suggests the main $s$-process contributions are small for this lighter element whose abundance may be set by the weak $s$-process from massive stars which also influence the abundances of $\alpha$- elements Mg and Si. Evidence of small inter-cluster differences for Eu in excess of that expected from the `solar system' main $s$-process suggest either a modification of that $s$-process or contributions from an $r$-process which vary from cluster-to-cluster. Abundance differences (Figure \ref{almg_heavy}) for Mg, Al, Si, Cu and Zn between the most and least heavy-element enriched clusters also suggest that the $s$-process contributors are not the sole contributor of chemical tags. For a given [Fe/H], it would appear -- and reasonably so -- that Type II and Type Ia supernovae have contributed in different proportions and possibly with a significant dispersion in relative abundances in their contributions.

The main $s$-process operates in thermally-pulsing AGB stars. In their He-shell, the relative  yields of Ba to Eu are set primarily by the neutron-capture cross-sections and, thus the fair match of cluster abundances to solar system $s$-process fractions is not surprising. Y and Zr are separated from Ba to Eu by the neutron magic number N=50 and $s$-process  abundances of Y and Zr relative to Ba to Eu are largely determined by the neutron flux in the He shell. 

Similar trends among the heavy elements are seen among field giants with [Fe/H] $\sim 0.0$, as anticipated. Consideration of heavy element abundances among thin disk field dwarfs also with [Fe/H] $\sim 0.0$ shows a spread in heavy element abundances. The age spread in the samples differ substantially: our OC range from 0.1 to 1.0 Gyr but the field dwarfs average about 6 Gyr. There is an indication that the heavy element chemical tag changes with time. Our present sample of OCs should be expanded to include OCs with ages greater than 1 Gyr. A few such OCs from both the thin and thick disks have been analysed for iron and lighter elements, as our literature survey \citep{Reddy2015,Reddy2016} showed, but data on heavy elements is essentially lacking. Comprehensive analyses of older OCs would provide a direct comparison with field stars of the thin and thick disks.
These analyses deserve to be matched with an exhaustive study of the thin and thick disks. Of especial note is the fact discussed in Section \ref{abu_diff:clusters}, presently available analyses of six associations all with [Fe/H]$\simeq 0.00$ suggest that they lack the spread in heavy element abundances seen across the sample of open clusters. If this is not a reflection of an unrepresentative sample of associations or subtle systematic errors, it poses an interesting question awaiting an answer.

A simple inference from the presence of the chemical tag based on heavy elements is that thermally-pulsing mass-losing AGB stars have to differing degrees polluted molecular clouds, the birthplaces of open clusters, whose composition is otherwise established primarily by massive stars and their supernovae, Type Ia supernovae and possibly merging neutron stars. Variations in contributions from these other sources of nucleosynthesis primarily affect relative abundances of lighter elements (say, Mg to Zn). Expansion of the chemical tag to some or all of these elements will likely require an additional improvement in accuracy with close attention to sources of systematic errors; Table \ref{sigmas_oc} shows that the observed dispersions for [El/H] and [El/Fe] for many of the Mg to Zn elements are comparable to the dispersions from the measurement uncertainties.

This uncovered trend among the abundances of heavy elements would seem to provide a basis for chemically tagging open clusters. The total spread in La and Ce abundances is about 0.5 dex among the OCs examined todate. Mean abundances for a cluster are determined to about 0.07$-$0.08 dex including uncertainties due to the stellar atmospheric parameters and equivalent width measurements. It would appear that the abundance spread among OCs is such that several clusters may have identical patterns of heavy element to within the measurement uncertainties. While this overlap may impair the specificity of a heavy element tag, a combination of this tag with one or more elements such as Mg, Al and Si and/or Cu and Zn (see Figure \ref{almg_heavy}) may yield a more incisive tag. Such a more-detailed exploration of chemical tagging in the forward direction is left for a future paper. Pursuit of chemical tagging in the reverse direction is left for hardier souls.

\vskip1ex 
{\bf Acknowledgements:}
We thank the anonymous referee for a thorough reading of this manuscript and for useful comments and suggestions which have improved the Paper.
We are grateful to the McDonald Observatory's Time Allocation Committee for granting us observing time for this project. DLL wishes to thank the Robert A. Welch Foundation of Houston, Texas for support through grant F-634.

\end{document}